\documentclass[%
 reprint,
 amsmath,amssymb,
 aps,
 prb
]{revtex4-2}

\usepackage{graphicx}
\usepackage{dcolumn}
\usepackage{bm}
\usepackage{color}
\usepackage[colorlinks=true,linkcolor=blue,allcolors=blue]{hyperref}

\begin{document}

\preprint{prb}

\title{Role of Goldstone mode in nonequilibrium insulator under DC electric field}

\author{Xi Chen}
\affiliation{Department of Physics, State University of New York at Buffalo, Buffalo, New York 14260, USA}
\author{Jong E. Han}
\email{jonghan@buffalo.edu} 
\affiliation{Department of Physics, State University of New York at Buffalo, Buffalo, New York 14260, USA}

\date{\today}
\begin{abstract} 
Measurements of resistive breakdown in electronic systems under a DC electric field have shown that the threshold fields for the insulator-to-metal transition are significantly lower than predicted by single-electron excitation scenarios, such as the Landau–Zener theory. In this work, we propose an alternate mechanism of destabilizing ordered insulators under a DC electric field by fluctuations of the order parameters through the Goldstone mode excitation. The low-energy bosonic excitations receive energy from accelerated electrons and thus destroy the spontaneous symmetry breaking. Using the Keldysh Green’s function formalism, we numerically confirm that the Goldstone mode remains well-defined in the nonequilibrium steady state, while its nonequilibrium excitations are sensitive to the electric field. The effective temperature of the Goldstone mode increases much more rapidly than the electronic effective temperature, with the bosonic threshold field significantly smaller than the electronic one, which suggests that collective phase dynamics may further reduce the transition field to the experimental range via a purely electronic mechanism.

\end{abstract}

\keywords{Goldstone mode, Nonequilibrium Green's functions, Charge density wave material, Insulator-metal transition}
\maketitle

\section{\label{sec:level1}Introduction}

Nonequilibrium phase transitions driven by an external electric field are among the central topics in modern condensed matter physics. The resistive switching (RS) phenomenon~\cite{Ridley1963}, abrupt change of electric resistivity by orders of magnitude when subjected to an electric field, remains a great challenge. RS has been widely observed in various materials, such as semiconductors, Mott insulators~\cite{janod,JSLee2015APR,stoliar2013universal}, and charge density wave (CDW) materials~\cite{chen2023charge} and the RS materials have drawn significant attention in recent years due to their potential applications in non-volatile memory devices~\cite{janod,JSLee2015APR,VaskivskyiNComm2016}, neuromorphic computing~\cite{del_valle,ChengPNAS2021}, and other electronic applications~\cite{janod,del_valle,JSLee2015APR,cario2010electric}. Despite abundant proposals of RS applications, however, the underlying mechanism of the RS phenomenon is still under debate, partly because a wide diversity of the system can exhibit RS with different microscopic mechanisms. One of the main challenges is the mismatch between the experimentally observed switching fields and theoretical predictions. In experiments, typical RS materials with an energy gap of order $1$ eV~\cite{janod} have the insulator-to-metal transition (IMT) at switching fields $1-10^4$ V/cm~\cite{janod,bardeenPT}, which correspond to the electrostatic energy of sub-meV within the unit cell and are very small compared to the energy gap. However, the conventional Landau-Zener (LZ) model, which was adopted in the early theories of resistive transition~\cite{ong1979,zener}, predicted  switching fields on the order of 
\begin{equation}
  E_{\rm LZ}\sim \frac{\Delta^2}{e\hbar v_F},
\end{equation}
with the insulating gap $\Delta$ and Fermi velocity $v_F$, which is typically on the order of $10^{3}$ kV/cm for typical electronic energy scales. 

Over last decades, various theories have been proposed, which can be broadly categorized into two scenarios: electronic versus thermal. The electronic scenario holds that it is the electric-field-accelerated electrons that trigger RS, which includes various ideas such as the LZ tunneling in Mott insulators~\cite{oka2003breakdown,oka2005ground,SugimotoPRB2008,IkedaPRB2024}, avalanche of impact ionization~\cite{guiot2013avalanche} and the multi-band Hubbard model~\cite{mazza2016field}. On the other hand, the thermal scenario argues that the Joule heating created by the electric current causes a rise of temperature and thus triggers the resistive transitions~\cite{zimmers2013role,nkumar2025,DiazPRB2023}. However, the theories developed so far cannot fully explain the apparent mismatch with the experimentally observed switching fields. 

Another important aspect of resistive switching is spatial inhomogeneity~\cite{JSLee2015APR,Lange2021PRA,Guenon2013EPL,li2017microscopic,YangNComm2012,AlspaughPRApp2026}, often considered crucial in driven systems. In realistic materials, the transition often involves the formation of metallic domains or filamentary conduction paths, which can further lower the switching field and modify the transport behavior. Out of equilibrium, the transitions are often strongly discontinuous and the pattern formation may partly account for the mismatch. While it has recently been pointed out~\cite{AlspaughPRApp2026} that switching fields can be reduced by several times due to the filament formation facilitated by disorder, the scale discrepancy is unlikely to be resolved by one mechanism. In the present work, therefore, we mainly focus on the microscopic mechanism on length scales shorter than those of domain formations~\cite{nkumar2025}.

Within the homogeneous description, previous studies of nonequilibrium phase transitions have clarified some important aspects of resistive switching. In particular, the single-electron nonequilibrium theory showed that the electronic and thermal scenarios can be combined through a field-dependent effective temperature, but the resulting switching field remains on the order of the energy-gap scale~\cite{han2018}. More recently, a quantum avalanche mechanism was proposed to explain the much lower switching field observed in correlated materials through phonon-assisted instability~\cite{han2023correlated,chen2024avalanche}. These developments suggest that the reduction of the switching field may be mediated by additional energy-absorption channels beyond direct single-particle excitation. This motivates us to examine whether collective electronic excitations can provide a new pathway under a DC electric field.

Motivated by this collective-excitation viewpoint, we focus on insulators formed by symmetry-breaking, such as the CDW, which constitute an important class of RS systems. In a simple one-dimensional picture, the ordered phase is characterized by a periodic modulation of the electronic density~\cite{gruner2018density,grunerRMP,khomskii2010basic,ZhuPNAS2015}. The low-energy physics can therefore be captured by a minimal two-band description, in which the energy gap arises from the order parameter. Its amplitude characterizes the magnitude of the charge density, while its phase specifies the positional shift of the density modulation. Therefore, symmetry-broken insulators naturally support two types of collective excitations: amplitude fluctuations and phase fluctuations~\cite{thorne1996charge,sugai2006phason}. While the transport properties and threshold-field behavior of CDW materials have often been discussed in terms of classical depinning theory~\cite{grunerRMP,fukuyama1978,lee1974conductivity,lee1979,fisherCDW}, understanding of nonequilibrium transitions through the quantum fluctuation of collective modes remains incomplete.

\begin{figure*}
  \centering
  \rotatebox{0}{\resizebox{6.6in}{!}{\includegraphics{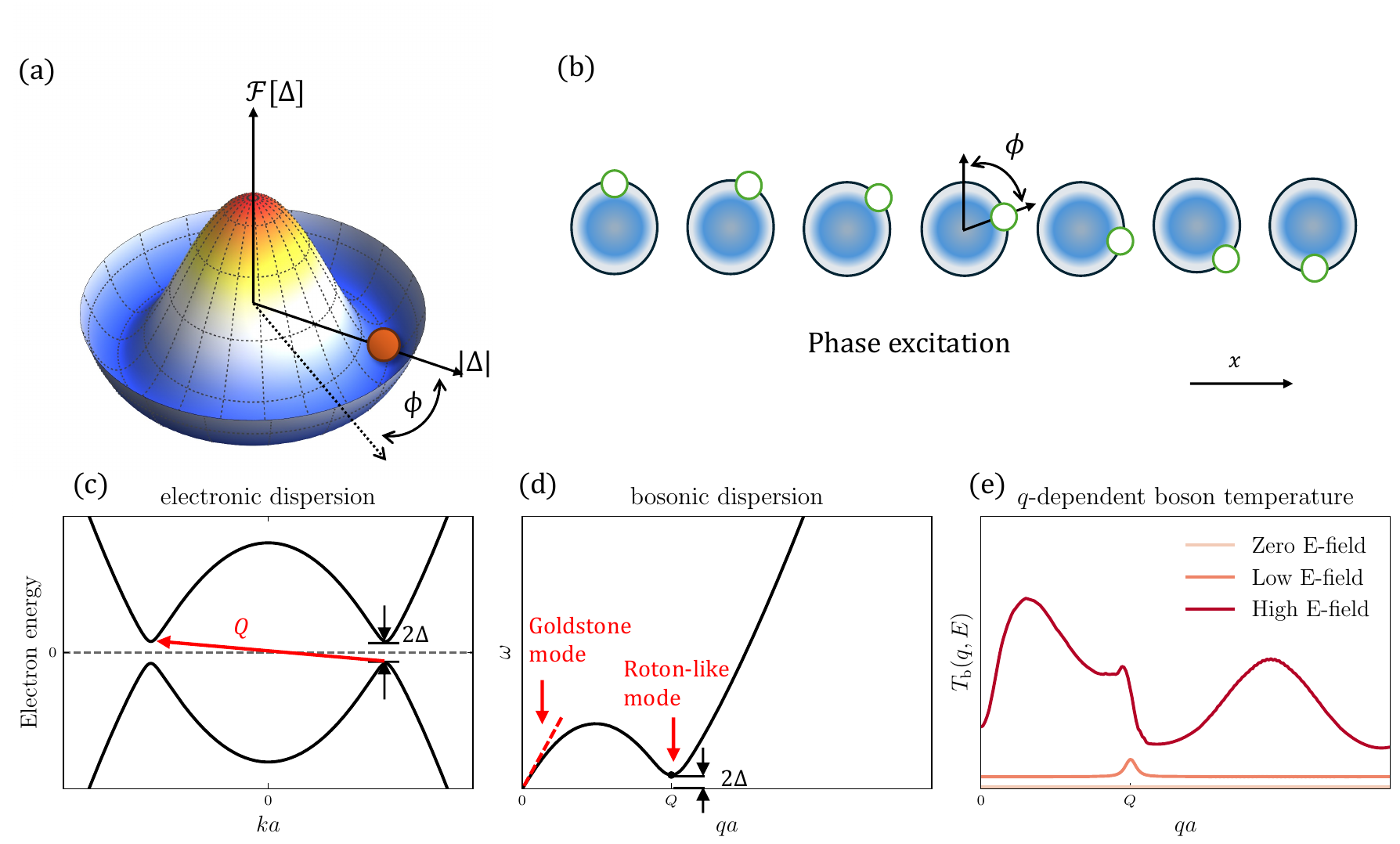}}}
  \caption{ (a) Schematic equilibrium free-energy landscape $\mathcal{F}[\Delta]$ for the complex order parameter $\Delta$. At low temperature the minimum occurs at a finite amplitude $|\Delta|$, forming a Mexican-hat-like profile with a continuum of degenerate minima: the free energy is insensitive to the global phase $\phi$, while the ground state spontaneously selects a specific $\phi$. (b) Illustration of a spatially varying phase texture $\phi(x)$ along the chain at fixed $|\Delta|$. Collective behavior of the phase corresponds to the gapless collective excitation (Goldstone mode) associated with the spontaneously broken symmetry. (c) Schematic electronic band structure of the symmetry-broken insulating state, where a gap $2\Delta$ opens around the Fermi level at the wavevector $Q$ connecting the two Fermi points. (d) Corresponding bosonic dispersion, showing a linear dispersion at small $q$ and a roton-like softening at finite momentum $Q$. (e) Schematic $q$-dependent boson effective temperature $T_b(q,E)$ under a DC electric field, illustrating that the small-$q$ modes are strongly excited at high fields.}
  \label{mexican}
\end{figure*}

This collective-mode structure can be understood more generally from the viewpoint of spontaneous symmetry breaking. For a symmetry-broken state described by a complex order parameter $\Delta=|\Delta|e^{i\phi}$, the Ginzburg-Landau free-energy can be represented by a Mexican-hat-like profile, as shown schematically in Fig.~\ref{mexican}(a). In the ordered phase, the system selects one minimum from a continuous manifold of degenerate states. Fluctuations around this minimum can be separated into radial and angular directions in the order-parameter space. The radial fluctuation corresponds to a massive collective excitation, often referred to as the Higgs mode, while the angular fluctuation corresponds to the Goldstone mode associated with the broken continuous symmetry~\cite{altland2010condensed,pekker2015amplitude,nagaosa2013quantum,tsuji2023higgs}. The distinction between these two modes is central to the mechanism considered in this work. The amplitude mode (Higgs mode) is gapped and therefore requires a finite energy scale to excite. In contrast, the phase mode (Goldstone mode) is gapless, making it much easier to excite under a nonequilibrium drive. Therefore, under a DC electric field, the phase degree of freedom may provide an efficient channel for absorbing energy from the field-accelerated electrons, beyond direct single-particle excitation across the full gap. Nevertheless, the behavior of the phase mode in nonequilibrium steady states under a DC electric field remains largely unexplored. In particular, it remains unclear how the phase mode interacts with nonequilibrium electron-hole excitations and whether it can provide a collective pathway for reducing the switching field.

The remainder of this paper is organized as follows. In Sec.~\ref{sec:level2}, we introduce the two-orbital lattice model describing the symmetry-broken insulating phase and formulate the problem within the nonequilibrium Keldysh Green’s function framework. We first discuss the equilibrium limit, establish the mean-field solution, and derive the collective phase mode, and then extend the formulation to nonequilibrium steady states under a DC electric field with dissipation. In Sec.~\ref{sec:level3}, we present numerical results for the electronic spectral functions, effective distribution functions, and polarization functions, and analyze the behavior of the Goldstone mode in nonequilibrium. In particular, we provide an analytic theory for the effective temperatures of electrons and collective modes and identify their respective threshold field behavior. Finally, in Sec.~\ref{sec:level4}, we summarize our findings and provide some discussions.

\section{\label{sec:level2}Model and formulation}

 We start from the Hubbard-like model in one dimension with two orbitals under a DC electric field. With the anticipation of a spontaneous gap formation, we start with a model of two bands crossing at the Fermi energy, such the zone-folding model in the example of the CDW. Whether the crossing bands remain gapless or become gapped will be decided self-consistently, as will be discussed below. For the gap-opening, the inter-band coupling is crucial. The intra-band interaction only contributes to the overall shift of the bands, unless it is very strong to drive a Mott transition. Therefore, we only consider the inter-band mixing as the interaction. For simplicity, we also ignore the spins. Then the lattice Hamiltonian reads as 
\begin{align}
  H_{\text{sys}} &= H_{\text{el}} + H_{\text{int}} , \nonumber \\
  H_{\text{el}} &=  t\sum_{j,\alpha=1,2}(-1)^{\alpha}(c^{\dagger}_{j,\alpha}c_{j+1,\alpha}+ h.c.) \label{model0} \\ 
& + \sum_{j,\alpha=1,2}(-1)^{\alpha}(\mu-2t) c^{\dagger}_{j,\alpha}c_{j,\alpha}, \nonumber \\ 
 H_{\text{int}} &=  -U \sum_{j} n_{j,1} n_{j,2}, 
\label{model}
\end{align}
where $c^{\dagger}_{j,\alpha} (c_{j,\alpha})$ is the electronic creation (annihilation) operator for orbital $\alpha(=1,2)$ (1 for the conduction band and 2 for the valence band) at site $j$ and $n_{j,\alpha} = c^{\dagger}_{j,\alpha}c_{j,\alpha}$. $t$ is the intra-band hopping parameter in the tight-binding model and $\mu$ is the chemical potential measured from the band edge. $U$ is the strength of the electronic interaction, and with the attractive interaction, this term can be considered as a generic mechanism for an insulator of broken symmetry.   

In the following discussions, we set the unit system $\hbar=k_B=e=1$. We set the hopping parameter $t$ with the discretization constant $a$ through $ta^2 =1$, for the unit of energy. In the following discussions, we compare numerical results against analytic calculations in the continuum limit, and we choose a small $a$ limit for numerical calculations with $a=0.4$, unless stated otherwise. Also, for comparison, we set $ta^2=\hbar^2/(2m)$ for the band mass $m$ in the continuum model.

Performing the Hubbard-Stratonovich (HS) transformation~\cite{altland2010condensed,hubbard1959calculation}, we can decouple the interaction term in Eq.~(\ref{model}) by introducing auxiliary fields, $\Delta_j$, and the interaction part of the Hamiltonian becomes
\begin{equation}
    H_{\text{int}} \Rightarrow H[\Delta] = \sum_{j} (\Delta_j c^{\dagger}_{j,1}c_{j,2} + h.c.)+ \sum_{j} \frac{|\Delta_j|^2}{2U}. 
    \label{interaction}
\end{equation}
The (complex) auxiliary field $\Delta_j$ fluctuates at each site and time independently.

In nonequilibrium steady-state problems driven by a DC field, a dissipation mechanism~\cite{han2013solution,han2013energy} should be introduced, where reservoirs are part of our total system in addition to Eq.~(\ref{model0}) and Eq.~(\ref{model}). However, for the sake of simplicity, we discuss the equilibrium limit in section~\ref{sec:level2-1} without dissipation, and will discuss dissipation in Section~\ref{sec:level2-2}.

As will be detailed below, our approach remains in the non-self-consistent random-phase-approximation (RPA) for the boson self-energy, and this work should be viewed as the first attempt for understanding nonequilibrium melting of the ordered insulator. Further implication of our results will be discussed in Section~\ref{sec:level4}.

\subsection{\label{sec:level2-1} Equilibrium limit }
In the zero-field limit, we approximate the Hamiltonian by imposing the static mean-field condition, and by suppressing the phase fluctuations, $\Delta_j(t)=\Delta$, which we assume to be real. By minimizing $\langle H[\Delta]\rangle$ in Eq.~(\ref{interaction}) with respect to $\Delta$, we obtain the mean-field condition
\begin{equation}
    \Delta = U \langle c^{\dagger}_{j,1}c_{j,2}+c^{\dagger}_{j,2}c_{j,1} \rangle. 
    \label{MFeq}
\end{equation}
$\Delta$ opens the gap between the two bands, as depicted in Fig.~\ref{mexican}(c). Introducing the field operator in the Nambu basis, $\Psi_{j}^{\dagger} = (c_{j,1}^{\dagger},\  c_{j,2}^{\dagger})$, and Fourier transforming to momentum space, the zero-field Hamiltonian becomes
\begin{equation} 
  H=\sum_{k}\Psi^{\dagger}_{k} \mathbf{h}_k \Psi_{k} + N\frac{\Delta^2}{2U},  
  \label{H2}
\end{equation}
where $\mathbf{h}_k$ is the matrix
\begin{equation}
  \mathbf{h}_k = \begin{pmatrix}
    \epsilon_k & \Delta \\
    \Delta & -\epsilon_k
  \end{pmatrix},
  \label{hk}
\end{equation}
where $\epsilon_k =2 t\cos(ka) + \mu -2t$. For notational simplicity, we denote $ka$ as dimensionless $k$ in analytic calculations. The effective free energy in equilibrium can then be computed as 
\begin{equation}
    \frac{\mathcal{F}}{N} = \frac{\Delta^2}{2U} - T\int_{\mathrm{BZ}} \frac{dk}{2\pi}\ln\left[ \cosh \left(\frac{E_{k}}{T}\right) \right] + \text{const.},
    \label{freeenergy}
\end{equation}
where $T$ is the temperature of the system and $E_k= \sqrt{\epsilon_k^{2} + \Delta^{2}}$ is the eigenvalue of Eq.~(\ref{hk}). The gap $\Delta$ can be regarded as the order parameter. The resulting free-energy is similar to that in the BCS theory~\cite{altland2010condensed}; in the present model, however, $\Delta=0$ corresponds to the metallic state, while finite $\Delta$ corresponds to the insulating state. The gap equation, Eq.~(\ref{MFeq}), can be evaluated by minimizing the free-energy as 
\begin{eqnarray}
    \frac{1}{U} &=& \int_{\mathrm{BZ}} \frac{dk}{2\pi} \frac{1}{E_k} \tanh\left(\frac{E_k}{T}\right)
    \label{gapeq} \\
    &\approx& \int_{\mathrm{BZ}} \frac{dk}{2\pi} \frac{1}{\sqrt{\epsilon_k^{2} + \Delta^{2}}} \quad (T\rightarrow 0).\nonumber 
\end{eqnarray}

To investigate the Goldstone mode, we introduce small fluctuations $\delta\Delta_j(t)$ in space and time around the static mean-field limit as $\Delta_j(t)=\Delta +\delta\Delta_j(t)$. Then the Hamiltonian can be separated into a mean-field part and the fluctuation part as~\cite{altland2010condensed,ZaikinPRL1997,vanOtterloEPJ1999}
\begin{eqnarray}
    H_{\Delta} &=& H_{\text{MF}} + H_{\text{fluc}}, \nonumber \\
    H_{\text{MF}} &=& \sum_{j} (\Delta c^{\dagger}_{j,1}c_{j,2} + h.c.) + \sum_{j} \frac{\Delta^2}{2U},  \\
    H_{\text{fluc}} &=& \sum_{j} (\delta\Delta_j c^{\dagger}_{j,1}c_{j,2}+\delta\Delta^*_j c^{\dagger}_{j,2}c_{j,1})
 + \sum_{j} \frac{|\Delta_j|^2-\Delta^2}{2U}.\nonumber 
    \label{interaction2}
\end{eqnarray}
Due to the Mexican-hat-like profile of the free-energy, Fig.~\ref{mexican}(a-b), we assume that the fluctuations mainly come from the phase and write the fluctuations in the form
\begin{equation}
    \delta\Delta_j \approx i\Delta\delta\phi_j,
\end{equation}
and the Hamiltonian for the fluctuation is governed by
\begin{equation}
    H[\delta\phi] = \sum_{j}\left[i\Delta(\delta\phi_j c^{\dagger}_{j,1}c_{j,2}-\delta\phi_j^* c^{\dagger}_{j,2}c_{j,1})
 + \frac{\Delta^2|\delta\phi_j|^2}{2U}\right].
\end{equation}

\begin{figure}
  \centering
  \includegraphics[width=\linewidth]{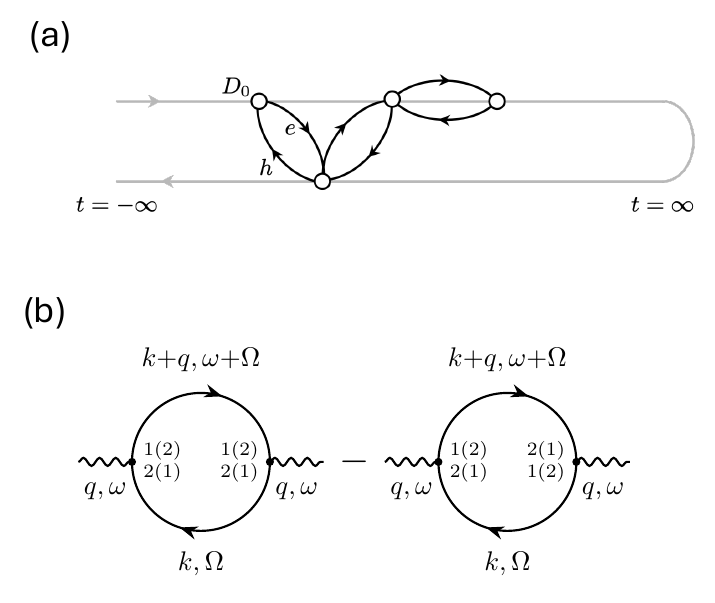}
  \caption{(a) Diagrammatic representation of the phase-mode propagation along the Keldysh time-contour (grey line) mediated by electron-hole excitations. The bare phase propagator (empty circle) introduced by the Hubbard-Stratonovich transformation is local in time. (b) The electron-hole bubble diagrams for the self-energy of the phase mode. The solid line is the electron GF with orbital indices $1$ and $2$, and the wavy line is the phase mode propagator.}
  \label{bubble}
\end{figure}

Having separated the mean-field and fluctuation parts of the interaction, we now employ the Keldysh Green's function (GF) formalism to analyze the phase fluctuation, so that the framework remains consistent with the subsequent nonequilibrium discussion. Because the Hubbard-Stratonovich transformation is performed locally in space and time, the auxiliary field is introduced as a local field variable. At the bare level, the phase fluctuation does not yet have its own spatial or temporal dynamics. Accordingly, its bare propagator is local in both space and time. The bare  GF of the phase field at momentum $q$ is local in time on the Keldysh contour as
\begin{equation}
    D_0(q,t) = -\frac{U}{\Delta^2}\delta(t), 
    \label{D0}
\end{equation}
and its Fourier transformation is $D^R_0(q,\omega) = -U/\Delta^2$. As depicted in Fig.~\ref{bubble}(a), the  phase mode acquires its nontrivial dynamics only through the motion of electron-hole pairs, and the full GF is obtained from the Dyson equation,
\begin{equation}
    D^R(q,\omega) = \frac{1}{D_0(q,\omega)^{-1}-\Pi_0^R(q,\omega)},
    \label{D}
\end{equation}
where $\Pi^R_0(q,\omega)$ is the retarded self-energy of the phase mode. It is obtained from the two types of electron-hole bubble diagrams shown in Fig.~\ref{bubble} as
\begin{align}
\Pi_0^R(q,\omega)
  &= \Delta^2\!\chi_0^{R}(q,\omega),
\label{selfenergy}
\end{align}
where $\chi^{R}_0(q,\omega)$ denotes the retarded electron-hole polarization function.

To compute the polarization functions in equilibrium, we start from the non-interacting retarded electron GFs $G^{R}(k,\omega)$ associated with the mean-field Hamiltonian $H_{\text{el}}+H_{\Delta}^{\text{MF}}$.
These retarded GFs are obtained directly by inverting the matrix $(\omega + i\eta)\mathbf{I} - \mathbf{h}_k$, where $\mathbf{h}_k$ is given in Eq.~(\ref{hk}) and $\eta$ is a positive infinitesimal number. The explicit form is
\begin{equation}
  G^R(k,\omega)
  = \frac{1}{(\omega + i\eta)^2 - E_k^2}
  \begin{pmatrix}
    \omega + \epsilon_k & \Delta \\
    \Delta & \omega - \epsilon_k
  \end{pmatrix} 
  \label{eqGRet}
\end{equation}
and the corresponding lesser- and greater- polarization functions are expressed as the convolution integral over momentum and frequency
\begin{align}
 \chi_0^{\lessgtr}(q,\omega)&=-i\sum_{k}{\sum_{\alpha\beta\alpha'\beta'}}^\prime\!\int\!\frac{d\Omega}{2\pi}(-1)^{\alpha+\alpha'} \nonumber \\
 &\times G^\lessgtr_{\alpha\alpha'}(k+q,\Omega+\omega)G^\gtrless_{\beta\beta'}(k,\Omega),
 \label{chiR_equil}
\end{align}
where the primed-sum over $\alpha\beta\alpha'\beta'$ is defined as $(\alpha,\beta)$ or $(\alpha',\beta')$ being chosen from the set $\{(1,2),(2,1)\}$, as suggested in Fig.~\ref{bubble}(b). The retarded polarization functions can then be computed from the lesser and greater ones by Hilbert transformation as 
\begin{equation}
    \chi_0^{R}(q,\omega) = \int_{-\infty}^{\infty} \frac{d\omega'}{2\pi} \frac{\chi_0^{>}(q,\omega') - \chi_0^{<}(q,\omega')}{\omega - \omega' + i\eta}.
    \label{HilbertT}
\end{equation}
With the straightforward calculation shown in Appendix.~\ref{app:1}, we can obtain the analytical expressions of the retarded polarization functions in the zero temperature limit as
\begin{align}
  &\chi_0^{R}(q,\omega)
  = \frac{1}{2}\int_{\mathrm{BZ}}\frac{dk}{2\pi}\Bigg[\left(1+\frac{\epsilon_k\epsilon_{k+q}+\Delta^2}{E_k E_{k+q}}\right) 
  \label{chi0} \\
  &\quad\times \left(\frac{1}{\omega + i\eta - E_{k+q} - E_k} + \frac{-1}{\omega + i\eta + E_{k+q} + E_k}\right)\Bigg].
  \nonumber
\end{align}

The dispersion relation of the Goldstone mode can be understood by analyzing the pole structure of the boson propagator $D^R(q,\omega)$, by rewriting as~\cite{ColemanBook} 
\begin{eqnarray}
   D^R(q,\omega)^{-1} & = & D_0(q,\omega)^{-1}-\Pi^R_0(q,\omega) \nonumber \\
   & = & D_0(0)^{-1}-\Delta^2\chi^R_0(0,0) \nonumber \\
   & &
   -\Delta^2[\chi^R_0(q,\omega)-\chi^R_0(0,0)].
\end{eqnarray}
The first line after the second equality is zero by the MF condition and the roots of the second line give us the dispersion relation of the phase mode. From Eqs.~(\ref{chi0}) and (\ref{gapeq}), we can easily verify $\chi_0^R(0,0)=-U^{-1}$ and $D_0(0)^{-1}-\Delta^2\chi^R_0(0,0)=0$, confirming the divergence of the Goldstone mode's response function in the $(q,\omega)\to 0$ limit~\cite{GoldstoneNuovo1961,GoldstonePR1962,mermin1966absence}. The Goldstone mode's dispersion relation can be obtained from the solution 
\begin{equation}
    \chi^R_0(q,\omega)-\chi^R_0(0,0)=0.
\end{equation} As shown in Fig.~\ref{mexican}(d), and later in Fig.~\ref{fchiRPA}, an explicit calculation of $\chi^R(q,\omega)$ confirms the Goldstone mode in the small-$q$ limit. The boson's lesser GF is computed as
\begin{equation} D^<(q,\omega)=\Delta^2\chi^<_0(q,\omega)|D^R(q,\omega)|^{2}.
\end{equation}

The limiting behavior of the dispersion relation of the phase mode for $q\to 0$ and $\omega\to 0$ can be analytically derived. As detailed in Appendix~\ref{app:1}, the real part of $\chi_0^R(q,\omega)$ is an even function of $q$ and $\omega$, respectively, and we can expand Eq.~(\ref{chi0}) to lowest order as
\begin{equation}
    \text{Re}\,\chi_0^{R}(q,\omega)-\text{Re}\,\chi_0^{R}(0,0)\approx A q^2 - B \omega^2,
    \label{gsdisp}
\end{equation}
with
\begin{equation}
    A = \frac{1}{4}\int_{\mathrm{BZ}}\frac{dk}{2\pi} \frac{(\partial_k \epsilon_k)^2}{E_k^3},\
   B = \frac{1}{4}\int_{\mathrm{BZ}}\frac{dk}{2\pi} \frac{1}{E_k^3}.
\end{equation}
Near the Fermi energy, $(\partial_k\epsilon_k)^2$ can be approximated as the Fermi velocity $v_F^2$
and the sound velocity of the Goldstone mode becomes $v_S=\omega/q=\sqrt{A/B}\approx v_F$.

By using Eqs.~(\ref{D}), (\ref{selfenergy}) and (\ref{HilbertT}), one obtains the Goldstone mode dispersion numerically, as shown schematically in Fig.~\ref{mexican}(d). In addition to the expected linear dispersion in the long-wavelength limit, the mode exhibits a roton-like minimum~\cite{shao2025electromagnetic,prange1990quantum} at a finite momentum $Q$. This momentum $Q$ is the wavevector at the band-crossing, as illustrated in Fig.~\ref{mexican}(c). At this wavevector, low-energy electron-hole (e-h) excitations are especially effective, enhancing the polarization response and softening the phase mode, thereby producing the roton-like minimum. As we show later, the effective-temperature analysis in nonequilibrium steady states supports this picture: the mode near the roton-like minimum heats up earlier under a DC electric field, as illustrated schematically in Fig.~\ref{mexican}(e). More detailed results are discussed in Sec.~\ref{sec:level3.3}.

\subsection{\label{sec:level2-2} Numerical formulation for nonequilibrium} 
We now extend the above equilibrium formulation to nonequilibrium steady states under a DC electric field~\cite{liprl2015}. To reach a nonequilibrium steady state, each site is coupled to a fermionic bath that can absorb excess energy created by the electric field. The total Hamiltonian then reads
\begin{equation}
    H_{\text{tot}} = H_{\text{el}} + H_{\Delta}^{\text{MF}} +H_{\text{E}} + H_{\text{bath}} + H_{\text{coup}},
\end{equation}
where the bath Hamiltonian $H_{\rm bath}$, its coupling to the main electronic chain $H_{\rm coup}$, and the driving term due to a DC electric field $H_{\rm E}$ are added to the previous Hamiltonian. The bath and electric-field Hamiltonians are
\begin{align}
    H_{\text{bath}} &= \sum_{j,\alpha=1,2}\sum_{n} (\epsilon_n - Ex_j )d^{\dagger}_{j,\alpha,n} d_{j,\alpha,n}, \\
      H_{\text{E}} &=  - \sum_{j,\alpha=1,2}E x_{j}c^{\dagger}_{j,\alpha}c_{j,\alpha},
\end{align}
where $d^{\dagger}_{j,\alpha,n} (d_{j,\alpha,n})$ is the creation (annihilation) operator with continuum orbital index $n$ coupled to orbital $\alpha$ at site $j$. The lattice position at site $j$ is $x_j=ja$. The bath consists of a fermion chain of length $L$ with dispersion relation $\epsilon_n$ at each site. The electronic system is coupled to the bath through
\begin{equation}
   H_{\text{coup}} = -\frac{g}{\sqrt{L}}\sum_{j,\alpha=1,2}\sum_{n} (c^{\dagger}_{j,\alpha} d_{j,\alpha,n} + h.c.),
\end{equation}
where $g$ is the coupling strength between the system and the bath. The bath is assumed to be in thermal equilibrium with temperature $T_{\text{bath}}$ and chemical potential $\mu$. The effect of the bath coupling can be incorporated exactly into the electronic self-energy,
\begin{align}
    &\Sigma_{j}^{R}(\omega) = -i\Gamma,\quad \Sigma_{j}^{<}(\omega) = 2i\Gamma f_{0}(\omega+Ex_j),\nonumber\\
    & \Sigma_{j}^{>}(\omega) = -2i\Gamma [1-f_{0}(\omega+Ex_j)],
\end{align}
where $f_{0}(\omega)=(e^{\omega/T_{\text{bath}}}+1)^{-1}$ is the Fermi-Dirac distribution function, and the hybridization function is $\Gamma=~\pi g^2 L^{-1} \sum_{n}\delta(\omega-\epsilon_n)$ for a flat bath density of states~\cite{han2013solution} in the continuum limit $L\to\infty$. In what follows, $\Gamma$ is interpreted as the electronic dissipation rate induced by the coupling to the bath. With this bath self-energy, the electronic retarded and lesser GFs are computed, with the computational details~\cite{han2013solution,han2013energy,liprl2015,han2023correlated} also found in previous publications. Since the bosons are not directly subject to the electric field, but are only implicitly affected through the electrons, the computation of the boson propagators follows the same procedure as the equilibrium method, as described in the previous section.

With local approximations, such as the dynamical mean-field theory~\cite{GeorgesRMP1996,AokiRMP2014,liprl2015}, the self-energy is local and the lattice summations are performed as in an effective impurity model. However, in this work, we are interested in the $q$-dependence of the Goldstone mode and its excitations. Therefore, we need to compute off-site Green's functions on a finite lattice block to perform Fourier transformation to the wavenumber. To this end, we define our system as a finite lattice of length $2N+1$ and two semi-infinite chains~\cite{mozumdar2025spectral}. The computation with the finite lattice is demanding, and, therefore, in the following Section~\ref{sec:level3} we selectively use the finite-lattice calculations only when they are needed.

With the central block of sites $(-N,\cdots,N)$, we use the Keldysh Green's function method to compute the full electron GFs for $N_s = 2N+1$ sites embedded in an infinite atomic chain, with the central site labeled by $0$. The retarded Green's-function matrix has the general form
\begin{align}
  [\mathbf{G}^R(\omega)^{-1}]_{ll'} &= 
  \mathbf{M}(\omega+lEa)\delta_{ll'} - \mathbf{T}\delta_{l,l'\pm 1} \notag \\
  &  - \mathbf{T} \mathbf{F}^R_{-}(\omega-(N+1)Ea)\mathbf{T}\delta_{l,-N}\delta_{l',-N} \notag \\
  & - \mathbf{T} \mathbf{F}^R_{+}(\omega+(N+1)Ea)\mathbf{T}\delta_{l,N}\delta_{l',N},
\end{align}
where 
\begin{align}
  \mathbf{M}(\omega) &= \begin{pmatrix}
    \omega + i\Gamma  -\tilde{\mu} & -\Delta \\
    -\Delta & \omega + i\Gamma  +\tilde{\mu}
  \end{pmatrix} \\
    \mathbf{T} &= \begin{pmatrix}
    t & 0 \\
    0 & -t
  \end{pmatrix},
\end{align}
where the subscript $ll'$ denotes the site indices and $\tilde{\mu}=\mu-2t$. Here $[\mathbf{G}^R(\omega)]^{-1}_{ll'}$ represents the $2\times2$ matrix block in row $l$ and column $l'$, while $\mathbf{F}^R_{\pm}$ are the retarded GFs of the semi-infinite chains extending to the left ($-$) and right ($+$) of the central $2N+1$ sites. The explicit forms of $\mathbf{F}^R_{\pm}$ are obtained iteratively from
\begin{align}
  \big[\mathbf{F}^R_{\pm}(\omega)\big]^{-1} &= \mathbf{M}(\omega) - \mathbf{T} \mathbf{F}^R_{\pm}(\omega \pm Ea) \mathbf{T}.
\end{align}
The $2\times2$ lesser (greater) GF matrix block is then computed from
\begin{equation}
    [\mathbf{G}^{\lessgtr}(\omega)]_{ll'} = \sum_{k=-N}^{N}[\mathbf{G}^{R}(\omega)]_{lk} \mathbf{\Sigma}^{\lessgtr}_{k,tot}(\omega) \big[\mathbf{G}^{R}(\omega)\big]_{l'k}^{\dagger},
\end{equation}
where the lesser (greater) self-energy matrix is given by 
\begin{align}
   \mathbf{\Sigma}^{\lessgtr}_{k,tot}(\omega) &= \Sigma^{\lessgtr}_{k}(\omega)\mathbf{I}+\mathbf{T} \mathbf{F}^{\lessgtr}_{-}(\omega-(N+1)Ea)\mathbf{T}\delta_{k,-N} \nonumber \\
    &\quad+ \mathbf{T} \mathbf{F}^{\lessgtr}_{+}(\omega+(N+1)Ea)\mathbf{T}\delta_{k,N},
\end{align}
and $\mathbf{I}$ is the $2\times2$ identity matrix. Similarly, the lesser  (greater) GFs of the semi-infinite chains, $\mathbf{F}^{\lessgtr}_{\pm}$, satisfy the iterative relations
\begin{align}
  \mathbf{F}^{\lessgtr}_{\pm}(\omega) &= \mathbf{F}^R_{\pm}(\omega) \big[ \mathbf{\Sigma}^{\lessgtr}_0(\omega) +\mathbf{T} \mathbf{F}^{\lessgtr}_{\pm}(\omega\pm Ea )\mathbf{T}\big]\big[\mathbf{F}^R_{\pm}(\omega)\big]^{\dagger}, 
\end{align}

Finally, the polarization function associated with the electron-hole excitations in the nonequilibrium steady state is
\begin{align}
  \chi_0^{\lessgtr}(r=l-l',\omega)
   &= -i{\sum_{\alpha\beta\alpha'\beta'}}^\prime\int\frac{d\Omega}{2\pi}(-1)^{\alpha+\alpha'} 
   \label{chi0_nonequil} \\
  & \times [\mathbf{G}^{\lessgtr}_{\alpha\alpha'}(\Omega+\omega)]_{ll'}[\mathbf{G}^{\gtrless}_{\beta\beta'}(\Omega)]_{l'l},\nonumber
\end{align}
where $r$ is the distance between two sites, and $\alpha$ and $\beta$ again denote the orbital indices. In our numerical calculations, all physical quantities are first computed in real space, after which a Fourier transform yields the momentum-dependent polarization function $\chi_0^{\lessgtr}(q,\omega)$.

\section{\label{sec:level3}Results and Analysis}
\subsection{\label{sec:level3.1}Electron Spectral Function and Effective Distribution}

\begin{figure*}
  \centering
  \rotatebox{0}{\resizebox{6in}{!}{\includegraphics{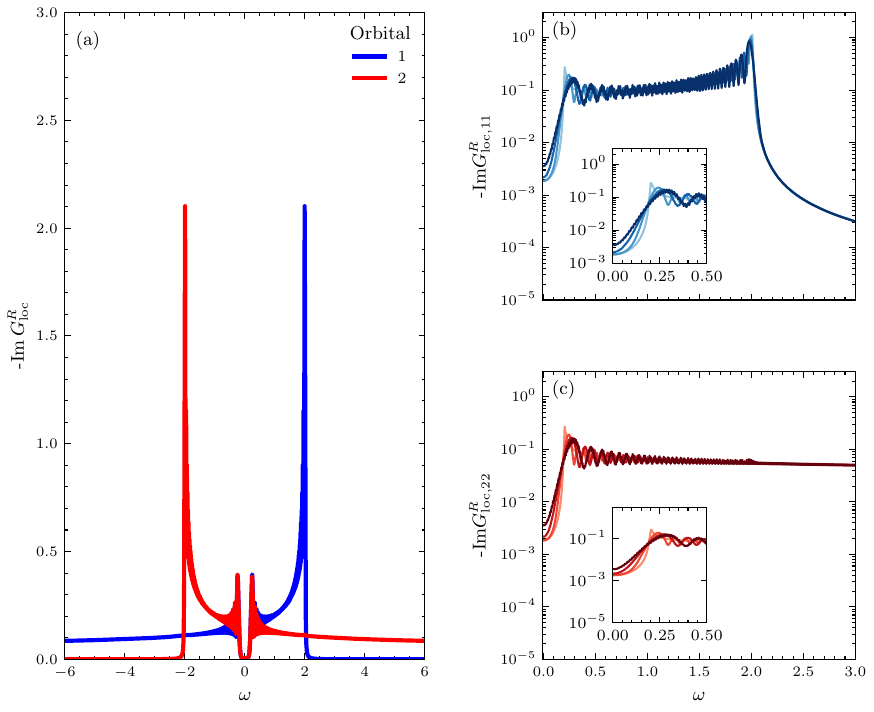}}}
  \caption{
(a) Local spectral function $-\mathrm{Im}\,G^{R}_{\mathrm{loc}}(\omega)$ for orbital~1 (blue) and orbital~2 (red) at
$E=0.005$, $\Gamma=0.005$, $\Delta=0.2$, $\mu=2$ (see the text for discussion of the tight-binding parameter $t$).
The bath temperature is fixed at $T_{\mathrm{bath}}=0.001$.
(b),(c) Local spectral functions for orbital~1 and orbital~2, respectively, for a set of DC fields
$E\in[0.001,0.006]$ with the step $0.001$ (from light to dark).
The insets highlight the gap region, where a finite in-gap spectral weight develops due to coupling to the dissipative bath.
}
  \label{espec}
\end{figure*}

We first present the numerical results for the local electron spectral function of each orbital at a nearly zero bath temperature, $T_{\text{bath}}=0.001$, obtained from the imaginary part of the local retarded GF, $-\mathrm{Im}\,G^R_{\text{loc}}$, as shown in Fig.~\ref{espec}. 
We choose the hopping parameter from $ta^2=1$ with the lattice constant $a=0.4$ (so that the kinetic energy takes the form as $k^2$ at small momentum $k$) and the chemical potential $\mu=2$ to have the Fermi level ($\omega=0$) near the band edge of the conduction and valence bands, as shown in Fig.~\ref{espec}(a). The form of Eq.~(\ref{model0}) ensures that the conduction and valence dispersion relations cross each other at the Fermi level, resulting in an insulating gap of size $2\Delta$.

Due to the dissipation medium coupled to the lattice, there is a small but finite spectral weight inside the gap $|\omega|<\Delta$ that may arise in realistic materials, such as impurities and coupling to substrates. Fig.~\ref{espec}(b) and (c) show the spectral evolution of the valence and conduction band, respectively, as the electric field $E$ is varied.

In this subsection, the analytic structures of the electronic Green's functions are discussed, complemented by numerical results, to prepare for understanding of the bosonic behaviors discussed in the next subsection. As has been shown earlier in Ref.~\cite{han2018}, the in-gap spectral weight at small fields can be well approximated by the zero-field limit. Near the chemical potential, the density of states is approximated by the Fermi velocity as $1/v_0\approx 1/(2ta^2)=1/2$.
While the spectral function increases gradually under an applied electric field inside the gap, the spectral function is well represented by the value at the Fermi level for $|\omega|\ll\Delta$,
\begin{equation}
  -\mathrm{Im} G^{R}_{\text{loc}}(\omega) \approx \sum_{k} \frac{\Gamma}{(v_0k)^2 + \Delta^2+ \Gamma^2} = \frac{\Gamma}{2 v_0 \Delta}.
  \label{gapspec}
\end{equation}
Therefore, the spectral function inside the gap increases linearly with $\Gamma$ and decreases with $\Delta$~\cite{han2018}.

In the in-band region, $|\omega|>\Delta$, the spectral function becomes oscillatory due to the Wannier-Stark effect, but its overall magnitude is not significantly modified compared to the zero-field case, as shown in Fig.~\ref{espec}(b-c). Therefore, we approximate the in-band spectral function by the zero-field spectral function at $|\omega| > \Delta$. Away from the spectral peak from the band edge, the in-band spectra are well approximated for $|\omega|>\Delta$ as 
$
  -\mathrm{Im} G^{R}_{\text{loc}}(\omega) \approx 1/(2v_0).
  \label{bandspec}
$
Combining with Eq.~(\ref{gapspec}), we have the electronic spectral function $A(\omega)$ as
\begin{equation}
  A(\omega) = \frac{\Gamma}{2\pi v_0\Delta}\Theta(\Delta-|\omega|) + \frac{1}{2\pi v_0}\Theta(|\omega|-\Delta).
  \label{aspec}
\end{equation}

\begin{figure}
  \centering
  \rotatebox{0}{\resizebox{3.5in}{!}{\includegraphics{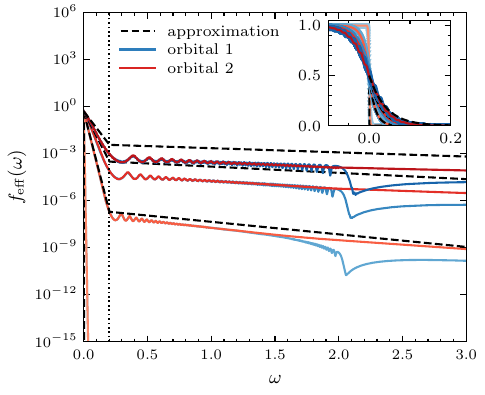}}}
  \caption{
    Nonequilibrium distribution of electrons, $f_{\rm eff}(\omega)$, on orbital $1$(blue) and $2$(red) at DC electric fields from $E=0$ (light color) to $E=0.006$ (dark) in steps of $0.002$. At $E=0$, $f_{\rm eff}(\omega)$ reduces to the equilibrium Fermi-Dirac distribution. With increasing $E$, the Fermi-Dirac distribution is smeared around the Fermi energy, reflecting field-induced excitations. The black dashed lines show the analytic approximation of Eq.~(\ref{feffapprox3}). The vertical dotted line marks the band edge $\omega=\Delta$. The inset plots the low-energy region on a linear scale to highlight the field-induced smearing near Fermi-level. The two orbitals exhibit nearly identical distribution, with only small difference near the band edge.
  }
  \label{fig:feff}
\end{figure}

We now discuss the nonequilibrium distribution for the electrons, with the numerical results shown in Fig.~\ref{fig:feff} for the electron distribution function for each orbital defined as
\begin{equation}
  f_{\text{eff}}(\omega) = -\frac{1}{2}\frac{\mathrm{Im} G^<_{\text{loc}}(\omega)}{\mathrm{Im} G^{R}_{\text{loc}}(\omega)}
  \label{feff}
\end{equation}
for diagonal GFs for $\alpha=1,2$. Except for the band-edge effect, the respective distribution functions are nearly the same. At zero field, the effective distribution function recovers the Fermi-Dirac distribution function. At finite fields, as shown in the log-scale plot of Fig.~\ref{fig:feff}, the frequency dependence shows two distinct exponential behaviors for $|\omega|<\Delta$ and $|\omega|>\Delta$.

The analytic behavior of $f_{\rm eff}(\omega)$ can be understood as follows. For $|\omega|>\Delta$, once an electron is excited into the upper band, its transport is metallic conduction with the mean-free-path $\ell=v_0/\Gamma$~\cite{mitra2008current,han2013energy} and the distribution function becomes the survival probability with the scattering rate $\Gamma$. For excitation of energy $\omega$, the particle has to survive the traveling distance of $\Delta  r=\omega/E$ with the probability $e^{-2\Gamma\Delta r/v_0}$. With the Landau-Zener probability $P_{\rm LZ}$~\cite{zener} for an electron to be placed in the conduction band, the distribution function for the in-band spectrum can be written as
\begin{equation}
      f_{\rm eff}(\omega>\Delta) \approx P_{\rm LZ}\, \frac{1}{2}e^{-2\Gamma\omega/(v_0 E)}.
  \label{feff_band}
\end{equation}

The in-gap distribution can be similarly understood~\cite{han2018}. The wavefunction attenuation through the gap is determined in a manner similar as that of the superconductivity theory~\cite{Tinkham1975Book}, where the coherence length is inversely proportional to the gap as $\ell\approx v_0/\Delta$. Therefore, the distribution function for $|\omega|<\Delta$ becomes  
\begin{equation}
      f_{\rm eff}(0<\omega<\Delta) = \frac{1}{2}e^{-2\Delta\omega/(v_0 E)}.
  \label{feff_gap}
\end{equation}
By combining the above results with the continuity at $\omega=\Delta$, we have the approximate analytic expression
\begin{align}
  f_{\rm eff}(\omega>0) &= \frac{1}{2}e^{-2\Delta\omega/(v_0 E)}\Theta(\Delta-\omega) 
  \label{feffapprox3} \\
  &+ \frac{1}{2}e^{-2\Delta^2/(v_0 E)}e^{-2\Gamma(\omega-\Delta)/(v_0 E)}\Theta(\omega-\Delta),\nonumber
\end{align}
where $\Theta(\omega)$ is the Heaviside step function. For $\omega<0$, $f_{\rm eff}(\omega)=1-f_{\rm eff}(-\omega)$. As shown in Fig.~\ref{fig:feff}, the analytic expression describes the numerical results quite well. 

\subsection{\label{sec:level3.2}Nonequilibrium Polarization Function and Goldstone Modes}

\begin{figure*}
  \centering
  \rotatebox{0}{\resizebox{6in}{!}{\includegraphics{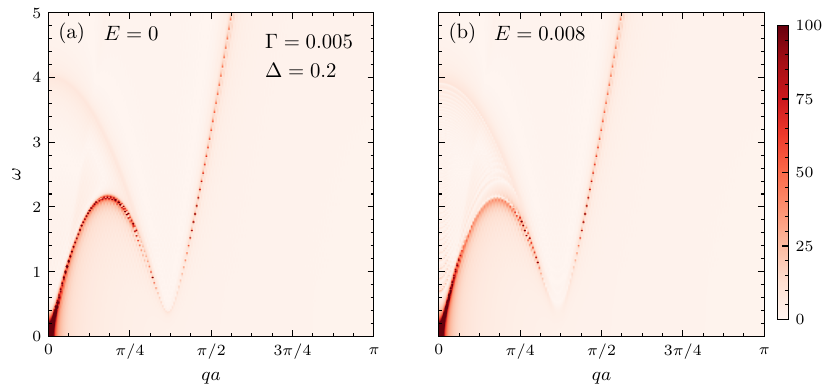}}}
  \caption{
    Color maps of the real part of the retarded boson propagator, $|\mathrm{Re}\,D^{R}(q,\omega)|$, used to identify the pole structure corresponding to the phase mode. Panels (a) and (b) show the numerical result at $E=0$ and $E=0.008$, respectively, at dissipation $\Gamma=0.005$, gap parameter $\Delta=0.2$ and lattice constant $a=0.4$. In both panels, well-defined dispersion relations confirm the existence of the Goldstone mode. At $E=0$ the linear dispersion relation agrees with the equilibrium benchmark; under the applied field the dispersion remains essentially unchanged, indicating that the Goldstone mode persists in the nonequilibrium steady state while the system remains gapped. 
  }
  \label{fchiRPA}
\end{figure*}

To verify the existence of the Goldstone mode in the nonequilibrium steady state, we use the numerical method described in Sec.~\ref{sec:level2-2} to compute the bare retarded polarization function $\chi^R_{0}(q,\omega)$ and then the bosonic GF, Eq.~(\ref{D}). In numerical calculations, we use a lattice model with the lattice constant $a=0.4$ and $N_s=201$ sites, for $q$-dependent GFs. 
Fig.~\ref{fchiRPA} shows the results for both zero and finite electric field. The pole structure in the retarded boson propagator is reflected in $|\text{Re}\,D^R(q,\omega)|$, directly revealing the dispersion relation and confirming the existence of the Goldstone mode. Despite being introduced as random Gaussian field that does not propagate in space and time,  $\Delta_j(t)$ acquires a particle-like property by coupling to electrons. In equilibrium ($E=0$), the mode is linear at small $q$ and exhibits a roton-like minimum at the band-crossing wavevector $Q$, fully consistent with the discussion above. This agreement serves as a benchmark for the numerical procedure. Under a DC electric field, the dispersion is essentially unchanged, indicating that the Goldstone mode survives in the nonequilibrium steady state until the gap collapses.

Having established the existence of the Goldstone mode in the nonequilibrium steady state, we next focus on the nonequilibrium behavior of the boson distribution through the lesser GF, $D^<(q,\omega)$. With $D_0(\omega)$ having no structure in $\omega$ [see Eq.~(\ref{D0})], the boson distribution inherits the statistical properties of the e-h polarization in $\chi^{\lessgtr}_0(q,\omega)$.  

\begin{figure}[h]
  \centering
  \rotatebox{0}{\resizebox{3.5in}{!}{\includegraphics{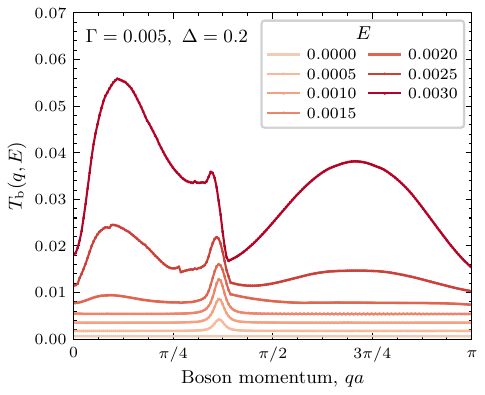}}}
  \caption{
Momentum-resolved effective temperature of the phase mode, $T_{\mathrm{b}}(q,E)$, extracted from the $q$-dependent boson distribution, Eq.~(\ref{Tbq}), in the nonequilibrium steady state, at $\Gamma=0.005$ and $\Delta=0.2$ for several DC electric fields $E$ (legend). With increasing field, $T_\mathrm{b}(q,E)$ becomes strongly momentum dependent, with the Goldstone modes at small $q$ leading the nonequilibrium excitations.
}
  \label{Teff_q}
\end{figure}

We show nonequilibrium excitations of bosons under a DC electric field in Fig.~\ref{Teff_q}. To quantify the nonequilibrium behavior, we represent the $q$-resolved excitations by integrating over the energy spectrum through the effective temperature, defined as~\cite{han2013energy}
\begin{equation}
   T_{\rm b}^2(q) = \frac{6}{\pi^2} \int_{0}^{\infty} \omega \frac{\mathrm{Im}\chi^<_0(q,\omega)}{2\mathrm{Im}\chi^R_{0}(q,\omega)} d\omega.
   \label{Tbq}
\end{equation}
For equilibrium $E=0$, this definition recovers the equilibrium temperature. As shown in Fig.~\ref{Teff_q}, the roton-like states at momentum $Q$ are excited early at small fields. As the electric field surpasses a certain value ($E\approx 0.0025$), the Goldstone modes receive dominant excitations. Despite the dispersion relation being almost unaffected by the electric field, the nonequilibrium excitations are strongly momentum-dependent and strongly affected by the field.

\subsection{\label{sec:level3.3}Local Approximation: Effective temperature of Bosons and electrons}

The discussions in the previous subsection demonstrated that the bosonic excitations are strongly $q$-dependent and are initiated by the Goldstone modes, which illustrates the fact that a nonequilibrium many-body theory may need to include non-local effects beyond the dynamical mean-field theory~\cite{GeorgesRMP1996,AokiRMP2014}. However, $q$-dependent calculations with a finite lattice with $N_s=201$ are quite demanding and suffer numerical noise at high fields, and we limit the following discussions to the local approximation $(N_s=1)$, which amounts to $q$-averaging.

\begin{figure*}
  \centering
  \rotatebox{0}{\resizebox{6in}{!}{\includegraphics{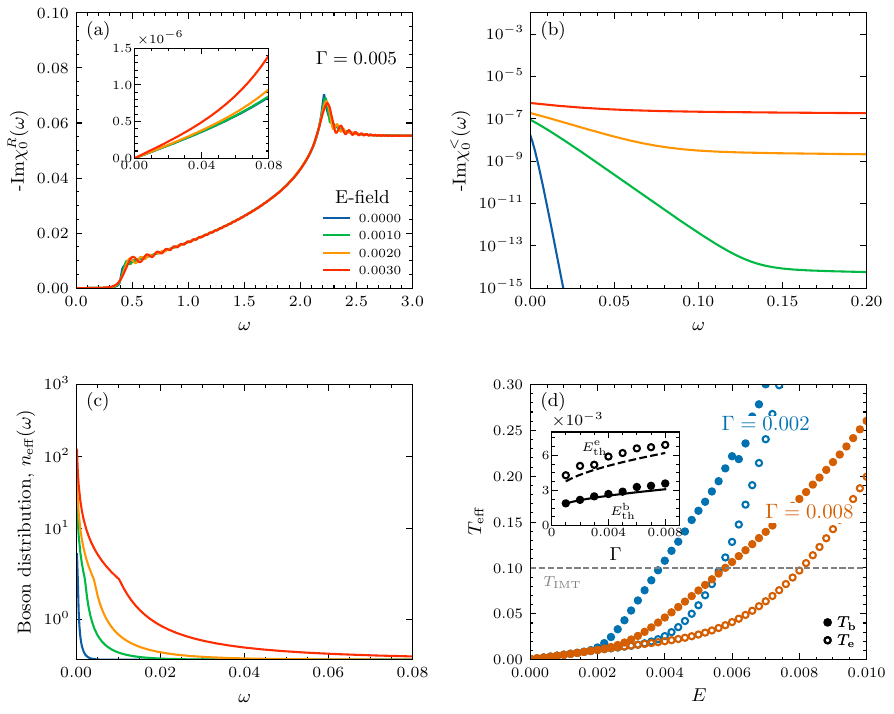}}}
  \caption{
Field-dependence of the local ($q$-averaged) electron-hole polarization spectrum, distribution and the resulting effective temperatures.
(a) $-\mathrm{Im}\,\chi^{R}_{0,\mathrm{loc}}(\omega)$, spectrum of the e-h states, showing the insensitivity to the DC field $E$; the inset zooms the
low-frequency region with the linear $\omega$-dependence.
(b) $-\mathrm{Im}\,\chi^{<}_{0,\mathrm{loc}}(\omega)$, the occupation spectrum of the e-h states, with strong dependence on $E$, in sharp contrast to (a).
(c) The corresponding boson (Goldstone-mode) distribution $n_\text{eff}(\omega)$, showing strong $E$-dependence with the excitations in the low $\omega (\ll\Delta)$ region. (d) Electrons effective temperatures $T_{\mathrm{e}}$ (unfilled) and Goldstone mode effective temperature $T_{\mathrm{b}}$ (filled) extracted from the local
effective distributions with different $\Gamma$; both temperatures follow the same small-field insulating-limit trend, but then exhibit a
rapid deviation at threshold fields, with $T_{\mathrm{b}}$ turning on at a lower field than $T_{\mathrm{e}}$. The numerical threshold fields $E_{\rm th}^{b}$ and $E_{\rm th}^{\rm e}$ are extracted from the intersections between the low-field linear behavior and the backward linear extrapolations of the rapidly increasing branches of $T_b$ and $T_{\rm e}$, respectively, and their dependence on $\Gamma$ is shown in the inset of panel (d) as symbols. The black solid and dashed lines, the analytical prediction for $E_{\rm th}^{\rm b}$ and $E_{\rm th}^{\rm e}$ from Eqs.~(\ref{bthresh}) and (\ref{ethresh}), respectively, show good agreement. The zero-field transition temperature $T_\text{IMT}$ (grey dashed line) is shown for comparison.
}
  \label{fig:boson}
\end{figure*}

We define the local effective distribution function of the Goldstone mode as
\begin{equation}
  n_{\text{eff}}(\omega) = \frac{1}{2}\frac{\mathrm{Im}\chi^{<}_{0,\text{loc}}(\omega)}{\mathrm{Im}\chi^{R}_{0,\text{loc}}(\omega)},
\end{equation}
where, as discussed above, we have used the relation $\mathrm{Im}D^{<}_{\text{loc}}(\omega)/\mathrm{Im}D^{R}_{\text{loc}}(\omega) = \mathrm{Im}\chi^{<}_{0,\text{loc}}(\omega)/\mathrm{Im}\chi^{R}_{0,\text{loc}}(\omega)$ which implies that the local Goldstone mode distribution is exactly equal to the electron-hole pair distribution. 
Fig.~\ref{fig:boson}(a-c) show the numerical results for $\mathrm{Im}\chi^{<}_{0,\text{loc}}(\omega)$, $\mathrm{Im}\chi^{R}_{0,\text{loc}}(\omega)$, and $n_{\text{eff}}(\omega)$. The spectral function of the Goldstone mode changes very little under an electric field, since the imaginary part of the retarded polarization function remains almost unchanged as the field varies. Note that $\mathrm{Im}\chi^{R}_{0,\text{loc}}(\omega)$ is the spectral function of the electron-hole pair and therefore displays a gap of order $2\Delta$. 
The occupation of the Goldstone mode exhibits a two-stage enhancement: the low-frequency sector $\omega \lesssim \Delta$ grows steadily with field, whereas the higher-frequency sector $\omega \gtrsim \Delta$ grows only slowly at small field but rises rapidly at larger field until it becomes comparable in magnitude to the low-frequency part where $\omega \lesssim \Delta$. 
This behavior implies that the Goldstone mode occupation can increase abruptly once the electric field exceeds a threshold, as discussed in Fig.~\ref{Teff_q}. 


Once the effective distribution functions of the electrons and Goldstone mode are obtained numerically, we define their local effective temperatures in the nonequilibrium steady state through~\cite{li2017microscopic}:
\begin{align}
  T_{\text{e}}^2 &= \frac{6}{\pi^2} \int_{-\infty}^{\infty} \omega \left[f_{\text{eff}}(\omega)-\Theta(-\omega)\right] d\omega, \label{elTeff}\\ 
  T_{\text{b}}^2 &= \frac{6}{\pi^2} \int_{0}^{\infty} \omega n_{\text{eff}}(\omega) d\omega,
  \label{gsTeff}
\end{align}
where $f_{\text{eff}}(\omega)$ is the total electron effective distribution function. The numerical results are shown in Fig.~\ref{fig:boson}(d). Both the electron effective temperature $T_{\text{e}}$ and the Goldstone mode effective temperature $T_{\text{b}}$ increase with electric field, and they share the same initial slope at small field, namely the insulating-limit behavior $T_{\text{e}}\sim v_0 E/\Delta$~\cite{han2018}, as suggested by Eq.~(\ref{feff_gap}). This agreement between the electronic and bosonic temperatures ensures that the theory obeys the linear regime out of the equilibrium fluctuation-dissipation limit.

As the field increases, the thermal equilibration between the electrons and bosons begins to break down. Interestingly, the boson temperature $T_{\rm b}$ begins to rise rapidly and deviates from the linear behavior earlier than the electron temperature $T_{\rm e}$. The existence of multiple effective temperatures in driven systems is well studied, such as the hot-electron and hot-phonon effects in semiconductors~\cite{Ferrybook}. Typically phonons play a passive role by receiving energy from the electrons as the main carrier of excess energy, and, therefore, the electron temperatures are higher than those of phonons~\cite{han2023correlated}. In this work, however, the electron-hole pairs receive energy directly from the electric field with the electron and hole getting accelerated individually. Due to the shallow excitation energy in the bosonic dispersion, the bosons become hotter earlier than electrons. While our approximation does not allow binding energy for an e-h pair, strongly bound e-h pairs may exhibit different thermal behavior, with the charge neutrality of the bound pair playing a more dominant role.

The inset of Fig.~\ref{fig:boson}(d) summarizes the behavior of the threshold electric fields, $E^\text{e}_\text{th}$ and $E^\text{b}_\text{th}$, for electron and boson, respectively, at which the respective effective temperatures deviate from the linear behavior. Overall, $E^\text{e,b}_\text{th}$ increases with the dissipation rate $\Gamma$, as a small $\Gamma$ tends to trap the nonequilibrium heat. The main result of this work is the observation
\begin{equation}   E_{\text{th}}^{\text{b}}/E_{\text{th}}^{\text{e}} \approx 0.5.
\label{ereduction}
\end{equation}

This reduction of the threshold electric field by the bosons has an important implication for the resistive switching. In the thermal mechanism of the resistive switching~\cite{li2017microscopic}, the nonequilibrium transition is well explained by the electron temperature matching the equilibrium transition temperature, $T_\text{e}\approx T_\text{IMT}\sim 0.5\Delta= 0.1$ [dashed horizontal line in Fig.~\ref{fig:boson}(d)], which leads to the estimate $E_\text{IMT}\sim 0.01$~\cite{han2018} without bosons. An explicit self-consistent calculation gives $E_\text{IMT}\approx 0.008$.


\subsection{\label{sec:level3.4}Analytic Understanding of Threshold Fields}

In the remaining part of Section~\ref{sec:level3}, we will provide an analytic justification of the relation, Eq.~(\ref{ereduction}). Understanding the electron temperature is quite straightforward. By combining Eqs.~(\ref{feffapprox3}) and (\ref{elTeff}), we have
\begin{equation}
T_{\text{e}}^2 \approx \frac{6}{\pi^2}\left(\frac{v_0E}{2\Delta}\right)^2 \left[1 + \left(\frac{\Delta}{\Gamma}\right)^2e^{-\frac{2\Delta^2}{v_0 E}}\right],
  \label{elTeffapprox}
\end{equation} 
where the approximation used are both $\Gamma \ll \Delta$ and the low-field limit. The second term in the bracket represents the nonequilibrium excitations due to the Landau-Zener transition. The threshold behavior occurs when the Landau-Zener excitations begin to dominate at $1\approx (\Delta/\Gamma)^2\exp(-2\Delta^2/v_0E^\text{e}_\text{th})$, or
\begin{equation}
    E^\text{e}_\text{th}\approx \frac{\Delta^2}{v_0\ln(\Delta/\Gamma)}.
    \label{ethresh}
\end{equation}
With $\Delta=0.2$, $v_0=2$ and $\Gamma=0.005$, this analytic expression gives $E^\text{e}_\text{th}\approx 0.0054$, which agrees with Fig.~\ref{fig:boson}(d). While $E_\text{IMT}\gtrsim E^\text{e}_\text{th}$ without bosons, Eq.~(\ref{ethresh}) can also serve as an analytic estimate of the IMT field~\cite{han2018}.

Understanding the boson nonequilibrium needs more work. As pointed out earlier, it is sufficient to examine the e-h GFs, $\chi^{R,<}_{0,\text{loc}}(\omega)$, to understand the boson thermal properties. The computation of the GFs starts with Eqs.~(\ref{C1}) and (\ref{C2}) in Appendix~\ref{app:3}. We first note from Fig.~\ref{fig:boson}(c) that the thermal spectrum is well confined for small $\omega\ll\Delta$. With this condition, we derive the nonequilibrium boson distribution function, as detailed in Appendix~\ref{app:3},
\begin{equation}
    n_\text{eff}(\omega)=
    \left(\frac{1}{4}+\frac{3v_0E}{8\Delta\omega}\right)e^{-\frac{2\Delta\omega}{v_0E}}
  +\frac{v_0E}{2\Gamma\omega}\left(\frac{\Delta}{\Gamma}\right)^2e^{-\frac{2\Gamma\omega}{v_0E}}e^{-\frac{2\Delta^2}{v_0E}}.
  \label{neff}
\end{equation}
In the manner similar as Eq.~(\ref{feffapprox3}), the first term is responsible for the linear regime and the second for the strong nonequilibrium excitations after the threshold field. Using the definition of Eq.~(\ref{gsTeff}), we obtain the effective temperature of the bosons as
\begin{equation}
T_{\text{b}}^2 \approx \frac{6}{\pi^2}\left(\frac{v_0E}{2\Delta}\right)^2 \left[1 + \left(\frac{\Delta}{\Gamma}\right)^4e^{-\frac{2\Delta^2}{v_0 E}}\right].
  \label{gsTeffapprox}
\end{equation}    
The first term in the square bracket is identical to the electron's $T_\text{e}^2$ and shows that electrons and bosons equilibrate in the linear regime up to the boson threshold field $E^\text{b}_\text{th}$. The main difference between electrons and bosons comes in the power to the factor $\Delta/\Gamma$. This leads to the boson threshold field
\begin{equation}
    E^\text{b}_\text{th}\approx \frac{\Delta^2}{2v_0\ln(\Delta/\Gamma)}
    \approx \frac12 E^\text{e}_\text{th},
    \label{bthresh}
\end{equation}
thus verifying the numerical result, Eq.~(\ref{ereduction}).

The crucial difference between the electron and boson temperatures is the exponent to the factor $\Delta/\Gamma$ — 2 and 4 for electrons and bosons, respectively, in Eqs.~(\ref{elTeffapprox}) and (\ref{gsTeffapprox}). The reason for the higher power comes from the fact that bosons are composite particles made of an electron at energy $\Omega$ and a hole at energy $\Omega+\omega$, with $\Omega$ anywhere on the single particle's spectrum. Therefore, even for $\omega\ll\Delta$, an e-h pair may have low-energy excitations from thermally excited e-h pairs with $\Omega>\Delta$. The single particle's spectral weight beyond the gap is enhanced by the factor $(\Delta/\Gamma)$ [see Eq.~(\ref{aspec})], and e-h pairs coming from the elevated levels $(\Omega,\Omega+\omega)$ have the additional enhancement factor of $(\Delta/\Gamma)^2$. Therefore, the nonequilibrium activation of the Goldstone mode is accelerated for small $\omega$, and this mechanism leads to the reduced threshold field $E^\text{b}_\text{th}$. 

By analytically understanding the mechanism of the phase mode excitations, we can speculate how it will impact the resistive switching beyond the approximation considered in this work. With the phase of the order parameter randomized by the field, the electronic gap will be smeared~\cite{WangPRB2023}, leading to a pseudogap-like spectrum in high-$T_C$ superconductors. With a softened gap, an electric field will be much more effective at exciting the Goldstone modes and will accelerate the melting of the global order. This destruction of the phase order will simultaneously reduce the gap amplitude in the mean-field condition, Eq.~(\ref{MFeq}). We expect that these combined effects will lead to a much more dramatic decrease of the insulating gap than suggested in Eq.~(\ref{ereduction}).

\section{\label{sec:level4}Discussion and Conclusion}

To understand the problem of the switching-field scale discrepancy in resistive switching, we investigated a mechanism for the role of Goldstone modes in symmetry-broken insulators. As an initial study of the problem, we explored the lowest-order effect of nonequilibrium excitations in the phase of order parameter, as the Goldstone modes, under a DC electric field. Despite the simple scheme of non-self-consistent random-phase-approximation (RPA) for both electrons and the Goldstone bosons, we demonstrated the possibility of an earlier resistive breakdown due to the nonequilibrium thermalization of the Goldstone mode.

By analyzing the pole structure of the GF of the phase mode within the RPA framework, we confirmed that the Goldstone mode remains well defined in the nonequilibrium steady state. 
We found, however, that the nonequilibrium excitation of the Goldstone mode is strongly enhanced and exhibits a two-stage enhancement. We also computed the effective temperatures of both the electrons and the Goldstone mode and found that both increase with electric field, but with different threshold behaviors. In particular, the effective temperature of the Goldstone modes begins to increase rapidly at a lower field than the electronic one, implying that collective phase fluctuations may reduce the insulator-to-metal transition field relative to a picture based only on single-electron excitations. 

Despite the confirmation that the Goldstone modes become hotter than the electrons and that the mechanism may provide a clue as to how to resolve the theoretical overestimation of the switching fields, the reduction predicted by Eq.~(\ref{ereduction}) remains insufficient. For comparison from experiments~\cite{janod,zimmers2013role}, transition-metal compounds, for example, with the band gap of $0.4$ eV (i.e., $\Delta=0.2$ eV) tend to have the switching fields of 10 kV/cm, which translates to $eEa=0.5$ meV if one uses the lattice constant of 5 \AA. Our estimate based on $T_\text{IMT}=T_\text{b}$ is roughly $eEa=$ 2-3 meV. Despite the significant improvement, the theory still seems a factor of $2\sim 10$ away.

The full effect of this mechanism can be achieved by considering self-consistency for both the electrons and the Goldstone mode, while we simultaneously update the mean-field condition, Eq.~(\ref{MFeq}). As demonstrated recently by Ref.~\cite{YangPRL2026}, the phase modes are excited more easily than discussed in this work when full fluctuation effects are included. Encouraged by this, we can replace the Green's functions for electrons and the bosons self-consistently, using a technique similar to that used in our previous work~\cite{han2023correlated,chen2024avalanche}. Furthermore, the electronic pseudogap formation due to the disordered phase~\cite{WangPRB2023} would expedite the collapse of the symmetry-broken insulators by a nonequilibrium drive.  This scenario could resolve a long-standing puzzle of the switching field discrepancy in correlated insulators in the future.

\begin{acknowledgments}
We acknowledge computational support from the CCR at University at Buffalo. We thank Piers Coleman for helpful discussions.
\end{acknowledgments}

\appendix

\section{\label{app:1} Analytical expression of the polarization function in equilibrium}
Inserting the equilibrium lesser and greater GFs obtained from Eq.~(\ref{eqGRet}) through the fluctuation-dissipation relation into Eq.~(\ref{chiR_equil}) and considering the Hilbert transformation shown in Eq.~(\ref{HilbertT}), the retarded polarization function can be expressed as
\begin{widetext}
  \begin{equation}
\chi_0^{R}(q,\omega) = \sum_{k}{\sum_{\alpha\beta\alpha'\beta'}}^\prime
\int_{-\infty}^{\infty} 
\frac{d\omega'}{2\pi}
\int_{-\infty}^{\infty}\frac{d\Omega}{2\pi} 
\frac{f(\Omega)-f(\omega'+\Omega)}{\omega - \omega' + i\eta}
(-1)^{\alpha+\alpha'}\mathrm{Im}G^{R}_{\alpha\alpha'}(k+q,\omega'+\Omega)\mathrm{Im}G^{R}_{\beta\beta'}(k,\Omega).
\label{chiR2}
\end{equation}
\end{widetext}
We first consider the normal bubble diagram $\chi_{0,\text{normal}}^{R}(q,\omega)$ for the retarded GFs with $\alpha=\alpha'$ and $\beta=\beta'$ in Eq.~(\ref{eqGRet}), and we obtain
\begin{widetext}
  \begin{align}
\chi_{0,\text{normal}}^{R}(q,\omega) 
&=\sum_k\Bigg[
(u_{k+q}^2v_k^2 + v_{k+q}^2u_k^2)\,
\frac{f(E_{k})-f(E_{k+q})}{\omega+E_k-E_{k+q}+i\eta}
+(u_{k+q}^2 u_k^2 + v_{k+q}^2 v_k^2)\,
\frac{f(E_{k+q})+f(E_k)-1}{\omega+E_k+E_{k+q}+i\eta}
\nonumber\\
&+(u_{k+q}^2 u_k^2 + v_{k+q}^2 v_k^2)\,
\frac{1-f(E_{k+q})-f(E_k)}{\omega-E_k-E_{k+q}+i\eta}
+(u_{k+q}^2v_k^2 + v_{k+q}^2u_k^2)\,
\frac{-f(E_k)+f(E_{k+q})}{\omega-E_k+E_{k+q}+i\eta}
\Bigg],
\label{chiRN}
\end{align}
\end{widetext}
where we introduce the coherence factors
\begin{equation}
u_k^2=\frac{1}{2}\left(1+\frac{\epsilon_k}{E_k}\right),\quad
v_k^2=\frac{1}{2}\left(1-\frac{\epsilon_k}{E_k}\right).
\end{equation}
Since $E_{k}=\sqrt{\epsilon_k^{2} + \Delta^{2}}>0$, in the low temperature limit $T\rightarrow 0$, the distribution function becomes $f(E_k)=0$, thus the Eq.~(\ref{chiRN}) simplifies to
\begin{align}
  &\chi_{0,\text{normal}}^{R}(q,\omega) 
=\sum_k
(u_{k+q}^2 u_k^2 + v_{k+q}^2 v_k^2) \nonumber \\
&\times\left(\frac{1}{\omega-E_k-E_{k+q}+i\eta} -\frac{1}{\omega+E_k+E_{k+q}+i\eta}\right).
\nonumber
\end{align}
Next, performing similar calculations for the off-diagonal elements with $\alpha\neq\alpha'$ and $\beta\neq\beta'$  from Eq.~(\ref{eqGRet}) in Eq.~(\ref{chiRN}), we obtain the abnormal bubble diagram $\chi_{0,\text{abnorm}}^{R}(q,\omega)$ as
\begin{align}
    &\chi_{0,\text{abnorm}}^{R}(q,\omega)= \sum_k(-2u_{k+q} u_k v_{k+q} v_k) \nonumber \\
&\times\left(\frac{1}{\omega-E_k-E_{k+q}+i\eta} -\frac{1}{\omega+E_k+E_{k+q}+i\eta}\right).
\nonumber
\end{align}
Combining the normal and abnormal diagrams and using an integral over $k$ to replace the discrete summation, we finally obtain
\begin{align}
&\chi_0^{R}(q,\omega)
  = \frac{1}{2}\int_{\mathrm{BZ}}\frac{dk}{2\pi}\Bigg[\left(1+\frac{\epsilon_k\epsilon_{k+q}+\Delta^2}{E_k E_{k+q}}\right) 
  \label{A6} \\
  &\times \left(\frac{1}{\omega + i\eta - E_{k+q} - E_k} + \frac{-1}{\omega + i\eta + E_{k+q} + E_k}\right)\Bigg].
  \nonumber
\end{align}

\section{\label{app:3} Calculation of the local polarization function in nonequilibrium steady states}
\begin{figure*}
  \centering
  \rotatebox{0}{\resizebox{6in}{!}{\includegraphics{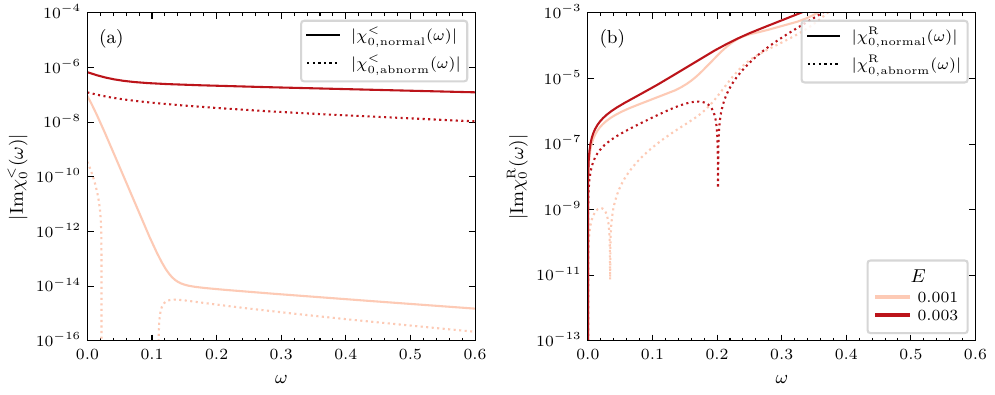}}}
  \caption{
  Comparison between the normal and abnormal electron-hole bubble diagrams, the first and second contribution in Fig.~\ref{bubble}(b), respectively at $\Gamma=0.005$, $\Delta=0.2$. 
  (a) The magnitude of the lesser component, $|\chi^<_{0,\mathrm{normal}}(\omega)|$ (solid) and $|\chi^<_{0,\mathrm{abnorm}}(\omega)|$ (dotted), plotted versus $\omega$ for two representative DC fields $E$ (light: $E=0.001$; dark: $E=0.003$).
  (b) The same for the retarded components, $|{\rm Im}\chi^{R}_{0}(\omega)|$.
  In both panels, the abnormal contributions are at least an order of magnitude smaller than the normal ones, and can be ignored in the analytic analysis.
  }
\label{appenchi}
\end{figure*}
To calculate the local effective temperature of the Goldstone mode, we compute the local lesser (greater) polarization function with the normal and abnormal bubble diagrams. However, the abnormal bubble diagram contribution is at least one order of magnitude smaller than the normal one, as demonstrated in Fig.~\ref{appenchi}, and, therefore, we only consider the normal bubble diagram. Thus, we utilize Eq.~(\ref{chi0_nonequil}) with the normal diagrams for $r=0$, and apply the Hilbert transformation given by Eq.~(\ref{HilbertT}) for the retarded polarization function. We then obtain the following expression of the local lesser (greater) and retarded polarization functions,
\begin{widetext}
\begin{align}
  -\mathrm{Im}\chi^{<}_{0,\text{loc}}(\omega) & \approx 2\pi^2\int_{-\infty}^{\infty}\frac{d\Omega}{2\pi}A(\omega+\Omega) A(\Omega)f_{\text{eff}}(\omega+\Omega)\left[1-f_{\text{eff}}(\Omega)\right], \label{C1}\\
  -2\mathrm{Im}\chi^{R}_{0,\text{loc}}(\omega) & \approx 2\pi^2\int_{-\infty}^{\infty} \frac{d\Omega}{2\pi} A(\omega+\Omega)A(\Omega)\left[f_{\text{eff}}(\Omega)-f_{\text{eff}}(\omega+\Omega)\right], \label{C2}
\end{align}
where we use the expressions of lesser and greater GFs, $G_{\rm loc}^{<}=2\pi i f_{\text{eff}}(\omega)A(\omega)$ and $G_{\rm loc}^{>}=-2\pi i [1-f_{\text{eff}}(\omega)]A(\omega)$. We used the symmetry between orbital 1 and 2, as shown in Figs.~\ref{espec} and \ref{fig:feff}, to drop the indices in $A(\omega)$ and $f_\text{eff}(\omega)$. 

For the lesser functions, the integrand of Eq.~(\ref{C1}) can be made an even function by shifting $\Omega$ to $\Omega-\omega/2$. In the limit of small $\omega\ll\Delta$, the integral can be approximated as
\begin{equation}
    \pi\left(\frac{\Gamma}{2 v_0\Delta}\right)^2 \int_{-\Delta+\omega/2}^{\Delta-\omega/2}d\Omega f_\text{eff}(\Omega+\omega/2)[1-f_\text{eff}(\Omega-\omega/2)]    +2\pi\left(\frac{1}{2 v_0}\right)^2\int_{\Delta+\omega/2}^\infty\frac{d\Omega}{\pi} f_\text{eff}(\Omega+\omega/2),
    \nonumber
\end{equation}
where the integral for $[\Delta-\omega/2,\Delta+\omega/2]$ is insignificant for small $\omega$. In the second integral, we used $1-f_\text{eff}(\Omega-\omega/2)\approx 1$. For a small field $E$, the upper integral limit can be replaced by $\infty$ in the first integral, and it becomes
\begin{equation}
    \frac{\pi}{4}\left(\omega+\frac{3v_0E}{2\Delta}\right)\left(\frac{\Gamma}{2 v_0\Delta}\right)^2e^{-\frac{2\Delta\omega}{v_0E}}.
    \nonumber
\end{equation}
The second integral becomes
\begin{equation}
    \pi\left(\frac{1}{2 v_0}\right)^2\frac{v_0E}{2\Gamma}e^{-\frac{2\Gamma\omega}{v_0E}}e^{-\frac{2\Delta(\Delta+\Gamma)}{v_0E}}.
    \nonumber
\end{equation}
Combining the results with the simplification $\Delta+\Gamma\approx \Delta$ in the above expression, we have
\begin{equation}
  -\mathrm{Im}\chi^{<}_{0,\text{loc}}(\omega) \approx 
  \pi\left(\frac{\Gamma}{2 v_0\Delta}\right)^2\left[
  \left(\frac{\omega}{4}+\frac{3v_0E}{8\Delta}\right)e^{-\frac{2\Delta\omega}{v_0E}}
  +\frac{v_0E}{2\Gamma}\left(\frac{\Delta}{\Gamma}\right)^2e^{-\frac{2\Gamma\omega}{v_0E}}e^{-\frac{2\Delta^2}{v_0E}}
  \right].
 \label{C5}
\end{equation}
\end{widetext}

Computation of the retarded function is much easier. We note in Eq.~(\ref{C2}) that, for $\omega\ll\Delta$, $f_\text{eff}(\Omega)-f_\text{eff}(\omega+\Omega)$ decays well inside the gap, and $f_\text{eff}(\Omega)-f_\text{eff}(\omega+\Omega)\ll f_\text{eff}(\omega+\Omega)[1-f_\text{eff}(\Omega)]$ for $\Omega>\Delta$ and the effects from the in-band states can be neglected, unlike in the lesser function. The integral is then approximated as
\begin{equation}
    \pi^2\int_{-\Delta}^\Delta\frac{d\Omega}{\pi}\left(\frac{\Gamma}{2 v_0 \Delta}\right)^2[f_\text{eff}(\Omega)-f_\text{eff}(\omega+\Omega)]
    \approx\pi\omega\left(\frac{\Gamma}{2 v_0 \Delta}\right)^2,\nonumber
\end{equation}
and obtain the retarded function
\begin{equation}
  -2\mathrm{Im}\chi^{R}_{0,\text{loc}}(\omega) \approx \pi\omega\left(\frac{\Gamma}{2 v_0 \Delta}\right)^2. 
  \label{chiret}
\end{equation}
With Eqs.~(\ref{C5}) and (\ref{chiret}), the nonequilibrium boson distribution $n_\text{eff}(\omega)$ becomes
\begin{equation}
    n_\text{eff}(\omega)=
    \left(\frac{1}{4}+\frac{3v_0E}{8\Delta\omega}\right)e^{-\frac{2\Delta\omega}{v_0E}}
  +\frac{v_0E}{2\Gamma\omega}\left(\frac{\Delta}{\Gamma}\right)^2e^{-\frac{2\Gamma\omega}{v_0E}}e^{-\frac{2\Delta^2}{v_0E}}.
  \nonumber
\end{equation}

\bibliography{reference}

\end{document}